# Statistical Study of Uncontrolled Geostationary Satellites Near an Unstable Equilibrium Point[◊]

Roberto Flores[*]
*Khalifa University of Science and Technology, P.O. Box 127788, Abu Dhabi, UAE*
*Centre Internacional de Mètodes Numèrics en Enginyeria (CIMNE), Gran Capità s/n, 08034 Barcelona, Spain*

Mauro Pontani[†]
*Sapienza University of Rome, via Salaria 851, 00138 Rome, Italy*

and
Elena Fantino[‡]
*Khalifa University of Science and Technology, P.O. Box 127788, Abu Dhabi, UAE*

The growth of the population of space debris in the geostationary ring and the resulting threat to active satellites require insight into the dynamics of uncontrolled objects in the region. A Monte Carlo simulation analyzed the sensitivity to initial conditions of the long-term evolution of geostationary spacecraft near an unstable point of the geopotential, where irregular behavior (e.g., transitions between long libration and continuous circulation) occurs. A statistical analysis unveiled sudden transitions from order to disorder, interspersed with intervals of smooth evolution. There is a periodicity of approximately half a century in the episodes of disorder, suggesting a connection with the precession of the orbital plane, due to Earth's oblateness and lunisolar perturbations. The third-degree harmonics of the geopotential also play a vital role. They introduce an asymmetry between the unstable equilibrium points, enabling the long libration mode. The unpredictability occurs just in a small fraction of the precession cycle, when the inclination is close to zero. A simplified model, including only gravity harmonics up to degree 3 and the Earth and Moon in circular coplanar orbits is capable of reproducing most features of the high-fidelity simulation.

---

[*] Senior Research Scientist, Department of Aerospace Engineering (Khalifa University); Affiliated Scientist, Transportation Group (CIMNE).
[†] Associate Professor, Department of Astronautical, Electrical, and Energy Engineering, Sapienza Università di Roma, Rome, Italy.
[‡] Associate Professor, Department of Aerospace Engineering, and Khalifa University Space and Planetary Science Center, and AIAA Permanent Member. Corresponding Author: elena.fantino@ku.ac.ae
[◊] *Preprint submitted to Journal of Guidance, Control & Dynamics on Jan 7, 2023*

## Nomenclature

| | | |
|---|---|---|
| $a$ | = | semi-major axis (km) |
| $A$ | = | cross sectional area (km$^2$) |
| $arcsec$ | = | arcsecond ($4.84814 \cdot 10^{-6}$ rad) |
| $dpy$ | = | degrees per year |
| $i$ | = | orbital inclination (rad) |
| $m$ | = | mass (kg) |
| $N$ | = | degree of expansion of the gravity field |
| $S$ | = | stable equilibrium points |
| $t$ | = | time (s) |
| $U$ | = | unstable equilibrium points |
| $\Delta \sigma$ | = | jump in longitude standard deviation (rad) |
| $\Delta t$ | = | duration of episode of sudden scatter increase (s) |
| $\Delta T$ | = | difference between TT and UT1 time standards (s) |
| $\{r, \theta, \lambda\}$ | = | geocentric spherical coordinates {radius, colatitude, longitude} (km, rad, rad) |
| $\sigma$ | = | standard deviation |
| $<x>$ | = | mean value of $x$ |

*Subscripts*
0    =  initial value

## I.    Introduction

The continuous expansion of the space debris population has become a major concern for the scientific community. Over the years, the Inter-Agency Space Debris Coordination Committee has published reports [1] and recommendations [2] to the interested bodies (companies and agencies) to mitigate the proliferation of space debris. The geostationary orbit is one of the most congested regions. While operators schedule disposal to a graveyard orbit for geostationary spacecraft at the end of operations, the maneuver is not always successful. Furthermore, old satellites were simply abandoned at their original station. Given the serious implications for the continued safe operation of active satellites, numerous algorithms for the planning and optimization of station-keeping maneuvers [3, 4], collision avoidance [5, 6], station change for on-orbit servicing [7] and active debris removal [8] have appeared in recent literature. Moreover, the long-term dynamics of decommissioned spacecraft in the geostationary orbit has become an area of active research (for example, see Refs. [9, 10]).

Two characteristic features of the orbital evolution of geostationary satellites are: (a) precession of the orbital plane and, (b) drift of the geographical longitude. Effect (a) is related to Moon and Sun gravitational perturbations, combined with Earth's oblateness. Early studies of this phenomenon are due to Allan and Cooke [11] and van der Ha [12], who employed double averaging (over the satellite orbital period and the periods of the perturbing bodies). Hechler [13] pointed out the occurrence of precession motion, with a period of 53 years, around an axis inclined 7.3 degrees from Earth's rotation axis. Friesen et al. [14] performed long-term propagations (up to 1000 years) of near-geostationary satellites and observed inclination changes of up to 15º over a cycle of 53 years. Recently, Proietti et al. derived simple analytical expressions for the evolution of inclination and right ascension of the ascending node of uncontrolled geostationary spacecraft [15].

Effect (b) is related to irregularities of Earth's gravity field. The $J_{22}$ harmonic of the geopotential, associated with the ellipticity of the terrestrial equator [16], plays a major role, due to resonance between Earth's rotation and orbital motion. It gives rise to two stable and two unstable equilibrium longitudes (for example, see Refs. [16, 17, 18]). The resulting orbital dynamics is either (i) librational or (ii) circulating. Separatrices divide these two qualitatively different behaviors. In case (i) the spacecraft oscillates around one of the stable longitudes, whereas in case (ii) it traverses all longitudes, moving either westward or eastward. With the action of $J_{22}$ alone, the longitudinal dynamics is completely predictable and depends on the initial conditions in a straightforward way. Lara and Elipe [19] proved the existence of stable periodic orbits emanating from the two unstable equilibria. However, $J_{22}$ alone is not sufficient to explain the complex longitude evolution of uncontrolled geostationary spacecraft. This was first noticed by Vashkoy'yak and Lidov [20], who pointed out the existence of librational motion spanning both stable points. They ascribed this phenomenon to higher-degree harmonics of the geopotential. Kiladze and Sochilina [21] and Kiladze et al. [22] confirmed this behavior, describing the occurrence of three types of dynamics depending on the initial conditions: simple libration around the closest stable position, long libration encompassing both stable positions, and circulation.

Kuznetsov and Kaizer [23] addressed the orbital motion of geostationary satellites located in the proximity of the separatrices. They tracked the regions where the separatrices migrate due to perturbations, and analyzed the dynamical behavior as a function of initial inclination and surface-to-mass ratio of the spacecraft. In a recent contribution, Celletti and Gales [9] employed Hamiltonian formalism and fast Lyapunov indicators to establish the amplitudes of the libration islands. Gachet et al. [24] modeled the orbital dynamics including solar radiation pressure and harmonics of the geopotential up to second degree. They focused on forced equilibrium solutions, proving they lie on a 5-dimensional torus. Colombo and Gkolias [10] investigated the long-term stability of the geostationary region for the purpose of designing effective disposal maneuvers.

A previous investigation of the complex longitudinal dynamics of decommissioned satellites [15] showed that initial positions sufficiently close to the unstable points trigger irregular dynamical behavior, with transitions between different types of motion (simple libration, long libration and circulation). This phenomenon requires harmonics of degree 3 of the geopotential, which are responsible for the asymmetry in energy of the two unstable points. The time-dependent third-body gravitational effects introduce a complex modulation of the total potential, allowing sporadic motion across unstable equilibrium points. Spectral analysis confirmed that the main contributors to this irregular behavior are gravity harmonics up to third degree and lunisolar perturbations. An interesting result of [15] was that small changes of the initial conditions can lead to vastly different longitudinal motion patterns. For some combinations of initial longitude and epoch (which influence the solution through the positions of Sun and Moon) the sensitivity to numerical perturbations is so extreme that it was impossible to obtain robust propagations beyond a horizon of 60 years. The observed phenomena are reminiscent of the complex structure underlying the dynamics of the GPS satellites, investigated by Daquin et al. [25] using analytical and semi-analytical techniques. Also, the analysis of the long-term evolution of some Molniya satellites (in 2:1 resonance with terrestrial rotation) has revealed the existence of a hyperbolic tangle in phase space and the lunisolar perturbation plays a key role in its structure [26].

This contribution seeks a deeper understanding of the phenomena observed in the geostationary regime, aiming to a better characterization of the sensitivity to initial conditions. It presents a Monte Carlo simulation of large collections of uncontrolled spacecraft released near one of the unstable equilibrium points. The evolution of the spatial coherence of the satellite cloud is characterized with a statistical analysis of the trajectories, identifying the physical effects responsible for the observed behavior. The main steps are: (a) propagation of the trajectories of a large collection of satellites, with initial positions randomly distributed near an unstable longitude; (b) statistical analysis of the trajectory set to quantify the spatial coherence and its long-term evolution; (c) identification of the relations between the statistical properties and the different perturbation sources; (d) use of simplified physical models to identify the dominant perturbations; (e) comparison of the simple models against high-fidelity numerical results to determine if they retain the qualitative dynamical behavior.

This paper is organized as follows. Section II deals with the physical model and the associated mathematical framework used for orbit propagation. Section III presents the numerical results obtained for a reference set of initial conditions. It focuses on the evolution of statistical parameters characterizing the distributions of geographical longitude and inclination. Section IV investigates the effect of initial conditions, i.e., reference epoch and initial longitude. Section V centers on simplified dynamical modeling, seeking to identify the physical effects that dominate the dynamical behavior. Finally, the main conclusions are drawn in section VI.

## II. Physical Model and Orbit Propagation

The trajectory is propagated in Cartesian coordinates using an adaptive embedded Runge-Kutta scheme of order 7(8) derived by Fehlberg [27]. The software has been validated against other propagators, and demonstrated the required level of accuracy [15]. The perturbations considered, following [15], are Moon and Sun third-body effects and solar radiation pressure.

While the satellite vector is formulated in the Earth-centered International Celestial Reference Frame (ICRF), the model of the terrestrial gravity field is expressed in the International Terrestrial Reference Frame (ITRF). The transform between the two frames follows the IAU 2000/2006 combined precession-nutation model [28]. Polar motion is neglected, because it cannot be reliably predicted into the future [29]. This introduces an uncertainty on the order of 0.3 arcsec in the calculation [29], but does not change the dynamical properties of the system. The spin of the Earth is computed according to [30], assuming $\Delta T$ increases 0.38 s per year (the approximate trend during the 2010-2020 period[*]). Instead of the complete IAU 2000/2006 precession-nutation model, a concise formulation based on Ref. [31] is used to reduce the computational burden while maintaining an accuracy of 1 arcsec.

The acceleration of gravity is obtained from a sum of spherical harmonics using the modified forward row recursion scheme [32]. Hereafter, whenever the expansion degree of the geopotential (*N*) is mentioned, it means that the geopotential model is complete to degree and order *N* (i.e., it includes all the relevant zonal, tesseral and sectorial harmonics). Gravity calculations conform strictly to the International Earth rotation and Reference Systems Services (IERS) recommended practice IERS Technical Note No. 36 [33]. It establishes EGM2008 [34] as the geopotential model of choice, and includes corrections for the secular drift of the low-degree zonal harmonics, as well as an improved value of the first zonal harmonic. The maximum degree of harmonic expansion in the calculations is $N = 8$, which yields the highest accuracy achievable with EGM2008 for a geostationary orbit [35].

---

[*] https://www.iers.org/IERS/EN/DataProducts/EarthOrientationData/eop.html

To minimize rounding errors in the calculation of the tidal forces of the Sun and Moon (third-body perturbations), the well-conditioned expression found in [36] is used. The positions of the celestial bodies are interpolated with cubic splines using tabulated state vectors from JPL's Solar System Dynamics website[*].

The acceleration due to radiation pressure is computed assuming the satellite is a sphere ("cannonball" model, see [37]) with 100% specular reflectivity. An area-to-mass ratio of $6.67 \cdot 10^{-3}$ m$^2$/kg, reasonable for a communications satellite, was used for the calculations. The eclipses were modeled with a cylindrical shadow approximation.

## III.    Numerical Results from Orbit Propagations

A previous study [15] suggested that the long-term evolution in longitude for uncontrolled satellites released near unstable equilibrium points is very difficult to predict. Some combinations of initial longitude and epoch were so sensitive to numerical perturbations (e.g., rounding errors) that it was not possible to obtain reliable solutions beyond 60 years into the future. While strongly hinting at the complexity of the long-term behavior, the original study was limited to two initial epochs, and widely-spaced (2 degrees apart) initial longitudes. To better characterize the unpredictability of the system, this work presents a detailed sensitivity analysis by means of a Monte Carlo simulation, propagating the trajectories of large clouds of spacecraft with very similar initial conditions. A statistical analysis of the results provides better insight into the long-term evolution towards disorder.

### A.  Choice of Initial Conditions

This work assumes that a spacecraft inside its operational window loses control abruptly. It is common practice among operators to maintain a window of +/- 0.05° centered on the nominal position. This is more stringent than International Telecommunication Union (ITU) requirements [38], which allow for a window twice as large. Furthermore, it is assumed that leaving the window during standard operations is a rare occurrence, considered a $3\sigma$ event. Thus, the standard deviation of the initial angular position (latitude and longitude) would be $\sigma_{\lambda,\theta} = 0.016°$ (57.6 arcsec). A binormally-distributed sample of longitudes and latitudes centered on the nominal position is generated using Marsaglia's polar scheme [39]. This sample provides the initial conditions for the trajectory propagation. In reality, the longitude and latitude of an active satellite are not independent random variables. Instead, they are controlled by the station-keeping strategy. However, the goal is to study the dynamical properties of the system irrespective of the peculiarities of any particular operator. In this respect, this approach is appropriate. Furthermore, it illustrates how the complex dynamics affect the normality of the initial distribution.

---

[*] https://ssd.jpl.nasa.gov/horizons.cgi

In a first approximation, the longitudinal dynamics of geostationary satellites can be described qualitatively by considering only harmonics of the gravitational field up to degree 2. The $J_{22}$ term, which represents the ellipticity of the equator, gives rise to four equilibrium points in the equator; two stable (75°E and 108°W) and two unstable (165°E and 15°W) – see [15] for a detailed derivation. Under the influence of $J_{22}$ alone, the motion of a satellite is a simple libration around the closest stable point. The third-degree sectorial and tesseral harmonics of the gravity field introduce an asymmetry in the potential of the unstable points [21]. Consequently, two additional modes of motion appear: "long" libration encompassing both stable points, and continuous circulation along the equator.

Additionally, Earth's polar flattening (quantified by the $J_2$ zonal harmonic) would cause the orbital plane to precess around the celestial pole. Third-body perturbations, on the other hand, force a precession around the pole of the orbit of the perturbing body. The combined effect of lunisolar perturbations and $J_2$ causes a 53-year precession cycle about the pole of the Laplacian plane, located between the pole of the ecliptic and the axis of rotation of the planet. It is inclined approximately 7º from Earth's axis of figure [11, 40] (this value is not really constant, due to the precession of the lunar orbit). As a result, the inclination of the spacecraft orbit varies between zero and 14º over the 53-year cycle. The interaction between the inclination cycle and the longitudinal dynamics (governed by the tesseral harmonics of the gravity field) modulates the potential barrier the spacecraft faces to move across the unstable points, enabling transitions between the different modes of motion [21]. This work focuses on trajectories starting close to the point of maximum instability, which offers the potential for highest complexity, enabling transitions between continuous circulation and long libration [15]. To this effect, 165.3ºE is chosen as nominal initial longitude for the sensitivity analysis. A survey of the literature reveals slightly different values for the position of the unstable point, depending on the model used to determine it (e.g., expansion degree of the gravity field). It is worth noting that, for a geostationary satellite, the order of magnitude of lunisolar perturbations is comparable to the irregularities of the gravity field [41]. Consequently, there is really no fixed equilibrium point, given that the relative orientations of Earth, Moon and Sun change continuously. Therefore, the exact value of the initial longitude is not critical as long as it is sufficiently close to the region of maximum instability. Due to the irregularities of the gravity field, the altitude where the centrifugal force acting on a geostationary satellite balances the inward acceleration due to gravity is different from the theoretical value for a spherical Earth. Using the theoretical height would induce a longitudinal drift right from the beginning of the computation. An iterative solver finds the appropriate height for each initial position in the cloud to ensure the initial motion is in sync with the rotation of the planet. The corrected starting position is approximately 600 m higher than the theoretical value for a spherical mass distribution.

Due to the importance of lunisolar perturbations, the initial epoch has a marked effect on the evolution of the cloud. The baseline propagation starts on 1 January 2020 at 0:00 UTC (JDN 2458849.5) because it is one of the dates used in Ref. [15]. In later sections, the effect of varying the initial epoch will be analyzed in detail.

**B. Longitude Evolution for the Baseline Case**

The reference solution includes a set of 600 trajectories starting on JDN 2458849.5, with the initial position bi-normally distributed around $<\lambda>$ = 165.3ºE and $<\theta>$ = 90º, and standard deviation $\sigma_{\lambda,\theta}$ = 0.016º. The length of propagation was set to 120 years, to encompass two complete cycles of precession of the orbital plane. The calculation considers harmonics of the gravity field up to degree and order 8, solar radiation pressure, lunisolar perturbation with precomputed ephemerides, and includes precession and nutation of Earth's axis. As indicated before, under the combined effects of lunisolar perturbations and Earth's polar flattening, the orbital inclination of a geosynchronous satellite will experience a 53-year cycle. Whenever the inclination is different from zero, seen from Earth the spacecraft will describe a figure eight trajectory (analemma) [42]. The North-South motion is due to the change in latitude along the orbit, while the East-West oscillation is caused by variations in the angular rate of motion of the projection of the spacecraft over the equator (the eastward velocity being minimal at the nodes and maximal at the points of extreme latitude). To isolate this diurnal modulation from the secular trend (which is the focus of the study), the trajectories have been sampled at an integer multiple of Earth's stellar period[*] (i.e., the rotational period in inertial space). This way, the snapshots of the trajectory are taken roughly at the same point of the diurnal cycle, effectively removing it from the output data. An output interval of 5 stellar days has been chosen, which provides a sufficient sampling of the lunar cycle.

---

[*] Not to be confused with the sidereal period, which is the time between passages of the reference meridian through the ascending node of the ecliptic. Due to the retrograde motion of the ascending node, the sidereal period is slightly shorter than the stellar day [43].

The longitudinal evolution of the satellite cloud (in terms of mean value and standard deviation) is depicted in Fig. 1. Note that, while the graph restricts longitude to the [0º, 360º[ interval, the calculations use a continuous representation (i.e., longitude is allowed to vary from -∞ to +∞) to compute the standard deviation. As far as position is concerned, values of the angular dispersion above 180º are not meaningful. Because two points in a circle can never be separated more than 180º, a large dispersion only indicates that the satellites are scattered all over the equator. However, the interest here is the dynamics of the system. The continuous representation longitude is used as an indicator of the accumulated East-West angular displacement. In that sense, it is relevant. For example, two spacecraft that move Eastwards 360º and 720º end up in the same longitude, but their dynamical behaviors are clearly different. The standard deviation of the continuous longitude remains sensitive to differences in drift rate, even when the spacecraft are uniformly distributed around the equator. Because the variations in scatter span six orders of magnitude Fig. 1 uses a logarithmic representation for the standard deviation. The most striking feature of the chart is the abrupt jump in scatter that takes place approximately half a century after the initial epoch ($t_0$). Initially, the longitudinal motion is relatively ordered ($\sigma_\lambda \sim 1º$) but, at the 50-year mark, a sudden transition to disorder takes place. 15 years later, the standard deviation reaches 250º (an increase of more than two orders of magnitude). This behavior is in stark contrast with the initial expectations of the authors, who anticipated a progressive evolution towards disorder characterized by an instability scale (i.e., exponential behavior). After the 65-year mark, the scatter grows monotonically (approximately 45º per year, the trend is very close to linear) until the next transition (see below).

There is a smaller transition 5 years after the initial epoch in Fig. 1 (the standard deviation changes only by a factor of five). Both occurrences (at 5 and 50 years) coincide with reversals of the direction of motion of the cloud. These reversals correspond to transitions between continuous circulation and long libration modes [15, 21], pointing strongly to a connection with the orbital plane precession cycle [40]. In fact, there is another change in trend, also accompanied by reversals, one century after the initial epoch. This is further evidence of a relation with the precession cycle, which has a period close to 50 years. Note that the 100-year transition takes place after the cloud has completely lost its spatial coherence (the standard deviation is above five revolutions). The state prior to this transition is so randomized that interpreting the results becomes difficult. Therefore, moving forward, the analysis shall prioritize the first two transitions (5 and 50 years after the initial epoch).

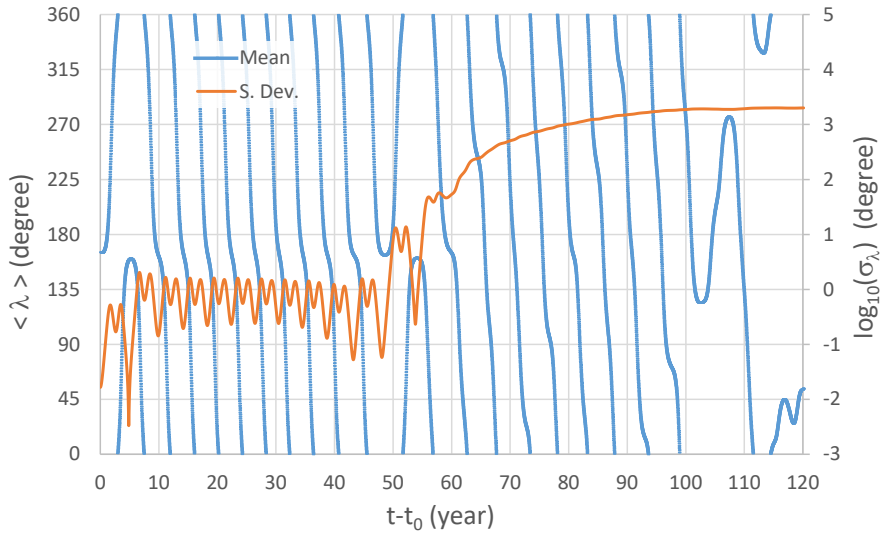

**Fig. 1 Longitude evolution for baseline configuration.**

A detail of the first 16 years of the simulation is shown in Fig. 2, to illustrate subtler features. The dashed horizontal lines in the chart indicate the stable (S) and unstable (U) equilibrium position, for reference. Besides the jump in scatter from 0.4º to 2º after five years, there is a cyclic variation of the standard deviation with a period of 4.2 years. It coincides with the time the cloud takes to circle the equator once.

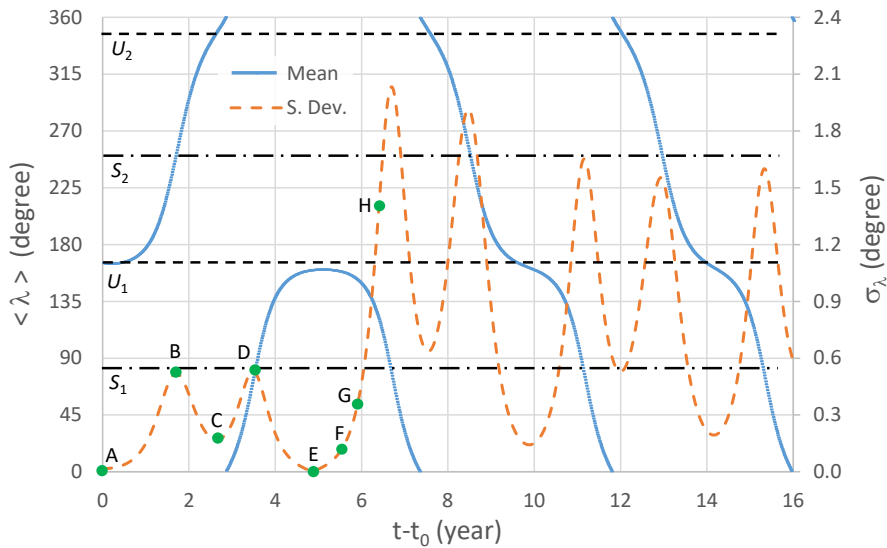

**Fig. 2 Longitude evolution for baseline configuration, detail for first 16 years.**

For each cycle, there are two maxima and two minima of the scatter. The maxima coincide with passages through the stable equilibrium longitudes 75°E/108°W ($S_1/S_2$ in the graph). When a spacecraft approaches a stable point, its longitude drift rate increases. Therefore, the leading satellites (those that arrive first to the equilibrium position) accelerate relative to those that lag behind. The end result is an increase in the spread of longitudes, giving rise to the maxima in the standard deviation curve. Conversely, when the cloud approaches an unstable point (165°E/15°W, $U_1/U_2$), the leading satellites slow down first, causing the rest to catch up and diminish the scatter. The asymmetry of the potential extrema, caused by the harmonics of the gravity field of degree 3 and higher [40], translates into differences in the magnitude of the peaks and valleys of the standard deviation. The highest maxima correspond to passages through $S_1$, while the deepest minima are associated with $U_1$.

Histograms of longitude help visualize the evolution of the cloud. The green circles with letters in Fig. 2 signal the times selected for plotting the histograms. Points A to E correspond to extrema of the standard deviation of longitude (A coinciding the initial epoch) which take place when the cloud crosses equilibrium points. Points E to H are evenly spaced at 6-month intervals to highlight the evolution after the direction of motion is reversed.

To improve the resolution of the histograms, they are computed with a cloud of 5000 satellites (using the same parameters for the random distribution of initial conditions). The mean value is subtracted from the longitudes to keep the curves centered on zero. Each histogram contains 300 bins of uniform size.

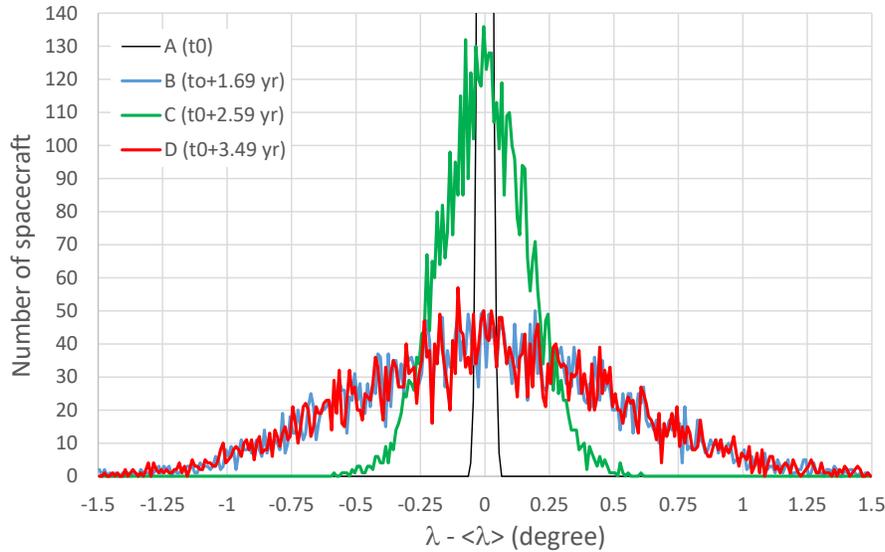

**Fig. 3  Longitude histograms for points A-D.**

The histograms for points A-D are displayed in Fig. 3. They illustrate the cycle of expansion-compression due to passages through the stable (curves B and D) and unstable (curves A and C) equilibrium points, respectively. The shape of the distribution does not change substantially, but the spread experiences large variations. The highest curve is clipped to reveal details of the others (the maximum value for histogram A is 1183). For points A to D, the changes in scatter, while substantial, are largely reversible. Thus, the motion of the cloud remain highly ordered. The behavior changes dramatically when the longitudinal drift is reversed (point E) as shown in Fig. 4. Again, curve E is clipped for clarity, because its maximum value is 2145[*]. The compression effect at E is remarkable: it reduces the standard deviation of longitude to 12 arsec, 80% smaller than the initial value (point A, 58 arcsec). This extreme squeezing does not preserve the order of the cloud; in fact, it causes a severe degradation. Once the motion is reversed (curves F to H), the shape of the distribution changes substantially (it becomes multi-modal) and the scatter experiences a rapid increase (see Fig. 2). This irreversible behavior is very different from that observed in Fig. 3. The reason is that, having started near the point of maximum instability, the spacecraft have enough energy to climb almost to the top of the potential barrier. Therefore, they can linger near the maximum for a comparatively long time before reversing course. However, the satellites do not start with the exact same energy (due to the random distribution of initial conditions). Furthermore, they do not reach the unstable point simultaneously, meaning they face slightly different potential barriers due to the continuously varying lunisolar perturbations. The net result is that those spacecraft that reach closer to the equilibrium condition (which, as explained before, is not even a fixed point in space) will slow down for longer. On the other hand, those with lower energy get deflected earlier and start to accelerate backwards first. Therefore, once all spacecraft have reversed course, the scatter of the cloud is increased substantially.

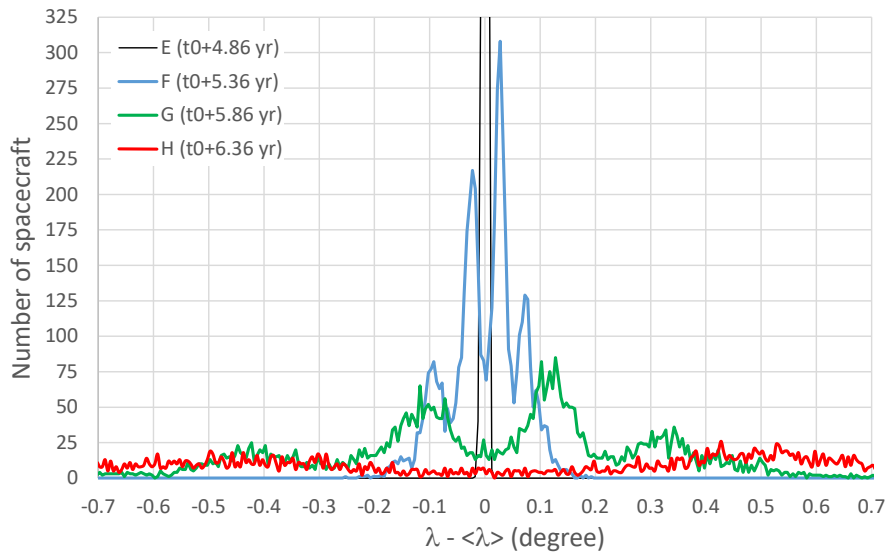

**Fig. 4 Longitude histograms for points E-H.**

---

[*] To realize the severity of the compression effect, note that the horizontal axis of Fig. 4 spans 1.4º vs. 3º for Fig. 3. Even using this narrower scale (i.e., smaller bins), the peak of curve E is higher than the maximum of A.

The transition 50 years after the initial epoch, which completely disrupts the order of the cloud, is driven by the same mechanism. The change in scatter is larger because the spread of the satellite cloud when it reaches the unstable point is significantly wider. Besides a greater range of spacecraft energies due to this spread, the lunar orbit experiences large variations over a 50 year interval, meaning the potential barrier to overcome is different from the one encountered at 5 years. This results in most spacecraft reversing course as before, but some are able to move across the unstable point. The motion of the cloud becomes incoherent, as there are satellites moving in both directions. This behavior is difficult to illustrate with longitude histograms, due to the extreme disorder involved. Fortunately, the statistics of the semi-major axis (Fig. 5) provide a clear, albeit indirect, depiction of the phenomenon.

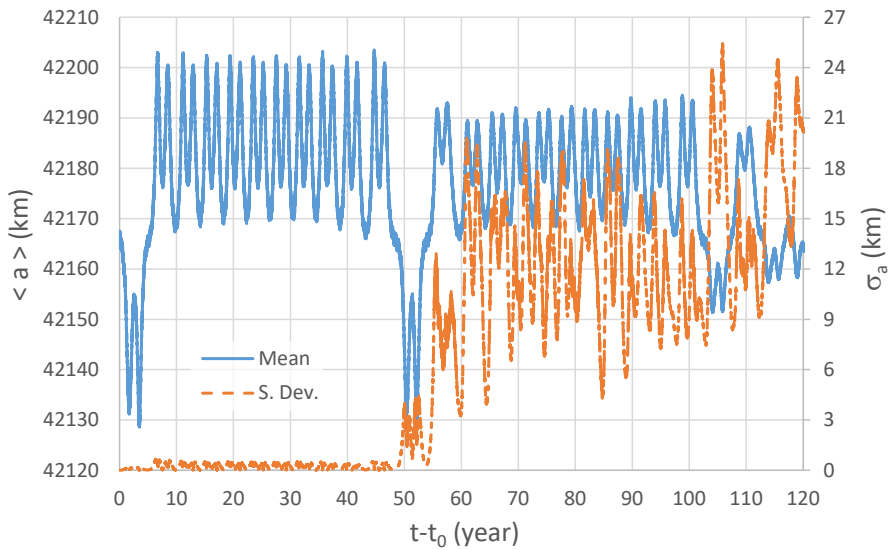

**Fig. 5  Semi-major axis evolution for baseline configuration.**

Satellites higher than the geosynchronous radius have orbital periods longer than one stellar day. This causes Earth rotation to overtake them and, seen from the ground, they drift West. Conversely, spacecraft at lower altitudes spin faster than the planet and move towards the East. Comparing Fig. 5 with Fig. 1, it becomes clear that the periods of eastward drift (at the start of the propagation and after 50 years) indeed coincide with the minima of semi-major axis. It is also noteworthy that the changes in altitude are comparatively small, tens of kilometers at most. Thus, moderate variations of height can translate into longitudinal motion reversals. Fig. 5 shows that, for the first 50 years, the standard deviation of the semi-major axis remains below 1 km, indicating that all spacecraft are moving in the same direction. Then, the scatter rapidly increases to tens of kilometers. This means there is a mix of satellites in high and low orbits, confirming that the direction of drift is no longer unique. These two populations of satellites move apart from each other continuously, causing the linear increase in standard deviation observed after 65 years.

## C. Inclination Evolution for the Baseline Case

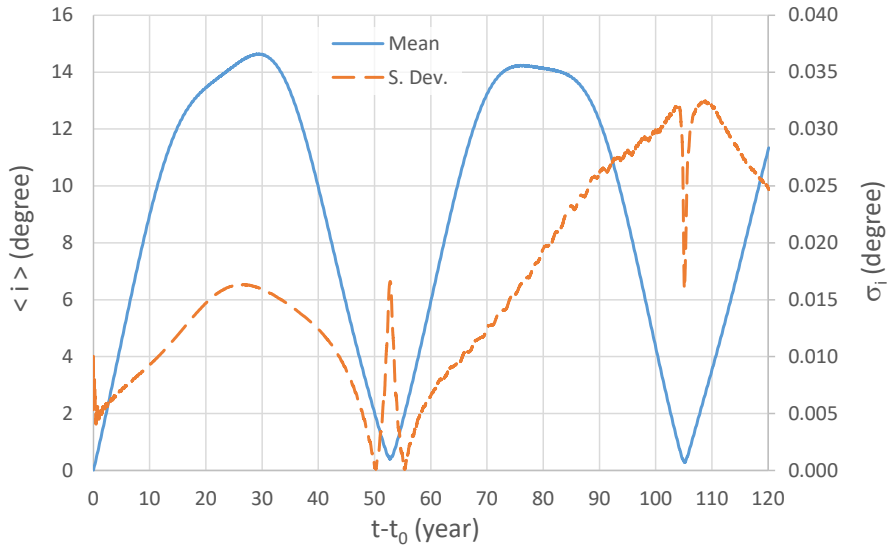

**Fig. 6  Inclination evolution for baseline configuration.**

Fig. 6 presents statistics for the orbital plane inclination. Comparison with shows that the episodes of longitudinal drift reversal and scatter increase take place near the minima of the mean inclination. The evolution of inclination follows closely the simplified analytical models (53-year cycle with maximum of 14º, [15]). The scatter is very small, remaining below 2 minutes of arc throughout the 120 years of simulation. The times of minimum inclination coincide with brief anomalies in the standard deviation. These anomalies are largely due to the definition of inclination. When the satellite orbits are very close to the equator, there will be positive and negative spacecraft latitudes at a given instant. However, the inclination of all orbits is positive (it is always inside the interval [0º,180º], by definition). This introduces an asymmetry in the distribution of inclinations, affecting the statistics. Nevertheless, it does not imply a change in physical behavior. For example, the initial standard deviation of inclination in Fig. 6 is 0.01º, smaller than the initial scatter in latitude (0.016º).

## D. Normality Evolution for the Baseline Case

This section explores how the shape of the initial distribution of longitudes and inclinations is affected by the evolution towards disorder. It also examines in more detail the effect of the apparent asymmetry introduced by the definition of inclination, using normality indicators (skewness and excess kurtosis) of the longitude and inclination (Fig. 7 and Fig. 8). Skewness characterizes the asymmetry of the distribution, while kurtosis is a measure of the importance of the tails (it increases with the presence of outliers). The $b_1$ estimator is used for sample skewness and the adjusted Fisher–Pearson standardized moment coefficient $G_2$ for kurtosis [44]. Both skewness and excess kurtosis should be close to zero for a representative sample drawn from a normal distribution. That is indeed the case for the initial longitude, as shown in in Fig. 7. On the other hand, the initial skewness of the orbital inclination is 0.98, due to the aforementioned asymmetry introduced by its definition. In fact, the skewness of a half-normal distribution[*] is approximately 1 [45], which agrees very well with the observed value. Fig. 8 shows that, over most of the first 70 years, both skewness and excess kurtosis of the inclination remain close to zero. This indicates that the spikes at the start of the simulation and after 50 years are mostly mathematical artifacts caused by the definition of inclination, and do not involve a physical change in the precession of the orbital planes.

Fig. 7 reveals a strong correlation between the longitudinal drift reversal episodes and the loss of normality of the longitude. There is a moderate spike in skewness and kurtosis 5 years after the initial epoch. This is followed by a period of relative stability where excess kurtosis stabilizes at -0.3 (coherent with the change from a normal to a multi-modal distribution, see Fig. 4) and skewness remains close to 0 (meaning the symmetry of the distribution is less affected). Finally, after 50 years, both parameters experience wild fluctuations and eventually stabilize at values quite different from zero, meaning even the symmetry of the distribution has been lost. Closer examination (right pane of Fig. 7) unveils a periodic modulation of skewness that spans the first 50 years of simulation.

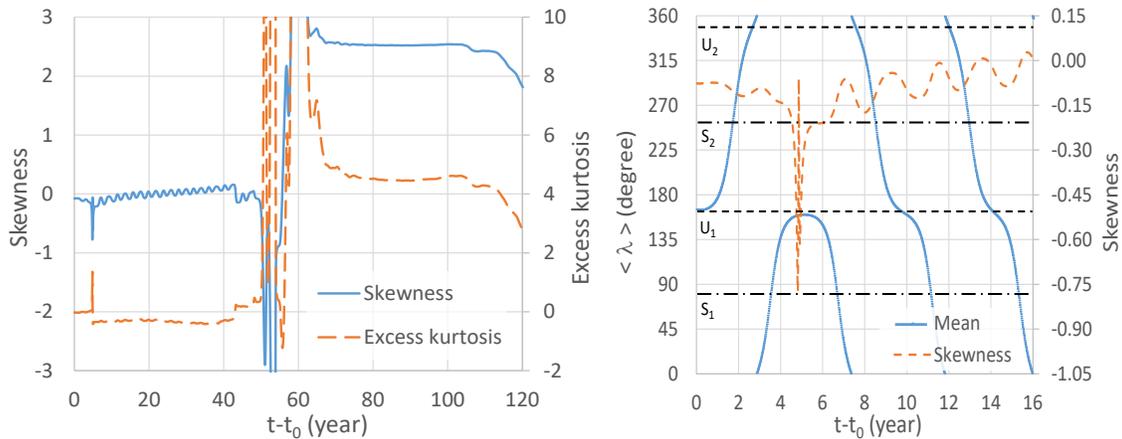

**Fig. 7 Normality indicators of longitude with detail for initial 16 years.**

---

[*] Given a normally-distributed variable of zero mean $x$, the probability density of $|x|$ is the half-normal distribution. That is, a zero-mean normal distribution is folded about the origin.

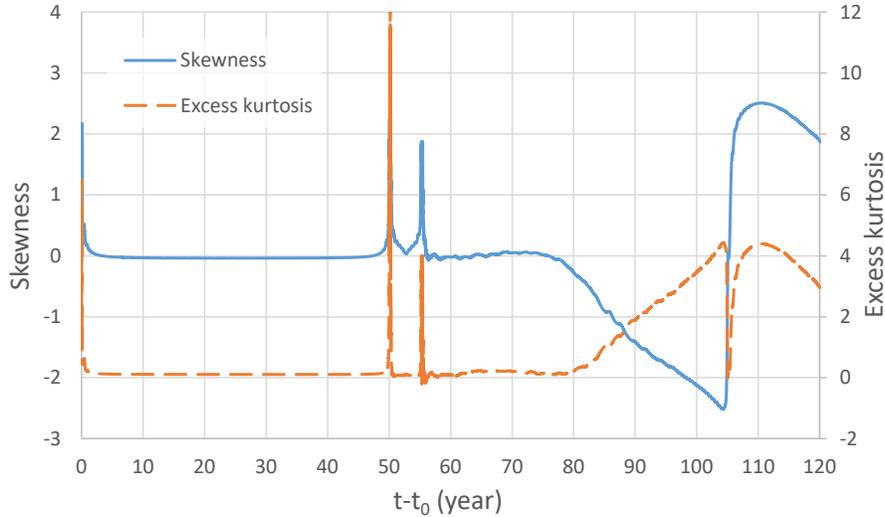

**Fig. 8 Normality indicators of orbital inclination.**

The skewness cycle reflects passages through $U_1$ and $U_2$. When the cloud of spacecraft approaches an unstable point, the leading satellites slow down first, compressing the forward tail of the distribution. Conversely, when the cloud leaves the equilibrium point, the leading spacecraft start to accelerate earlier. Thus, the forward tail stretches while the rear one remains squished. This alternating asymmetry translates into the skewness cycle. The spikes 5 years after the initial epoch are caused by the same compression-expansion effect, but exacerbated by the fact that the cloud comes to a complete stop before reversing direction (i.e., the squeezing is much more pronounced).

## IV. Effect of Initial Conditions

To achieve a good understanding of the system, it is imperative to characterize the influence of the initial conditions and try to eliminate those perturbation sources whose effect on the overall behavior is minor. This section starts with an overview of the effects of the initial longitude and epoch. Then, it removes layers of complexity from the physical model, to obtain the simplest formulation that retains the fundamental characteristics of the original system. This reduced model serves to narrow down the possible causes of the phenomenology observed.

**A. Effect of Initial Longitude**

From the simplified analytical formulations [40], transitions between continuous circulation and long libration (i.e., the pattern observed for the baseline case) are expected for initial longitudes close to the point of maximum instability (163ºE). Moving away from it, the long libration mode should gain preponderance, with the episodes of continuous circulation becoming shorter (because farther from the equilibrium point the initial energy is lower and it becomes more difficult to move across the potential barrier). For spacecraft starting far enough from the unstable point, continuous circulation is no longer possible, the motion becomes permanent long libration. These predictions agree well with the numerical simulations. For initial longitudes in the range 153ºE – 171ºE, the behavior is qualitatively similar to the baseline case, with the episodes of sudden disorder becoming less pronounced as the bounds of the interval are approached. For a more comprehensive review on the effect of initial longitude, see Ref. [15]. Note that the interval of longitudes where the complex behavior is possible changes with the positions of Sun and Moon, so it may be slightly different for other initial epochs. As an example, Fig. 9 displays the behavior for an initial longitude of 152ºE, just outside the interval of complex behavior. It shows continuous long libration without sudden increases in scatter. The cyclic variation of standard deviation due to passage across equilibrium points is present, like in the baseline configuration. On top of that cycle, there is a slow secular trend, with the average standard deviation over the cycle increasing by 2.5º over 120 yr.

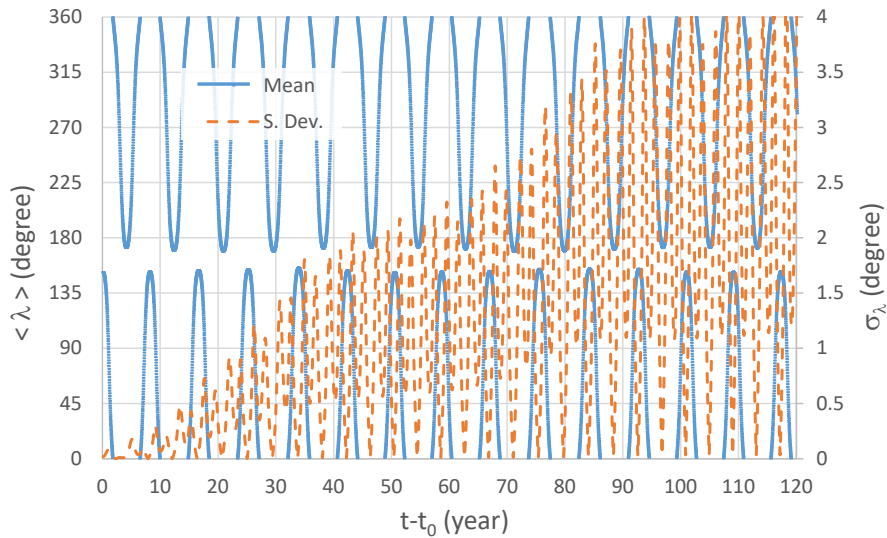

**Fig. 9 Longitude evolution: initial position 152ºE, initial epoch 1 Jan. 2020.**

**B. Effect of Initial Epoch**

To verify if the episodes of disorder are really linked to the orbital plane precession cycle, one can plot the time of occurrence of these episodes against the initial epoch of the simulation. Given that the satellites always start in equatorial orbits, the initial epoch coincides with the beginning of the 53-year inclination cycle. Therefore, the plot should be a straight line with unit slope. To build the plot, a date must be assigned to the transition. This requires defining an arbitrary convention because the episode is not instantaneous. In fact, there can be several consecutive reversals of the direction of motion over a period of a decade or longer.

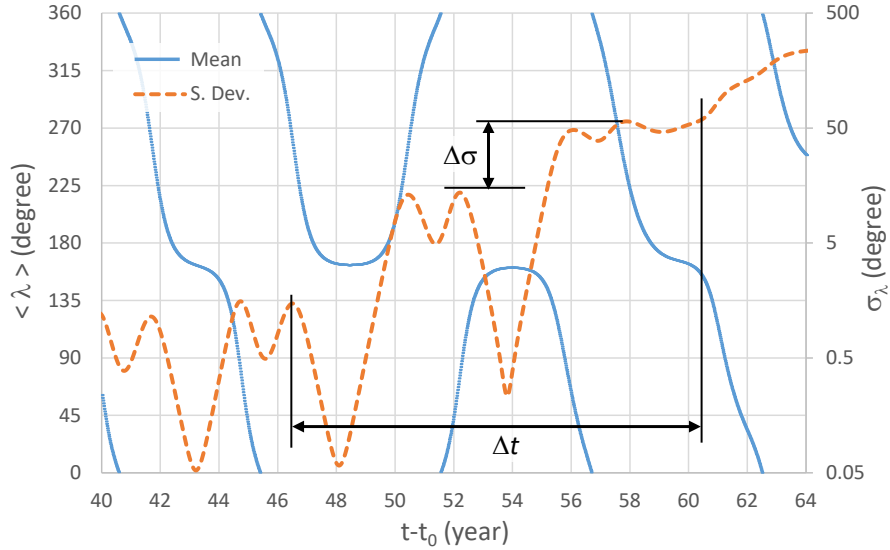

**Fig. 10  Convention for measuring duration ($\Delta t$) and strength ($\Delta \sigma$) of longitude jumps.**

Henceforth, it shall be assumed that the beginning and end of the transitions always take place at a local maximum of the standard deviation plot. The initial and final peaks are those that mark a change in the overall trend of the plot. As an example, Fig. 10 shows this criterion applied to the second transition of the baseline dataset. According to the rules stated above, the episode begins at the local maximum before the first jump in scatter ($t_0+46$ year) and finishes when the linear trend of the standard deviation starts ($t_0+60$ year). There is some degree of subjectivity in this choice, the point where the scatter starts to stabilize ($t_0+56$ year) could also be considered the end of the transition. What really matters is that the large variation in dispersion occurs in the neighborhood of 54 years after the initial epoch. In that respect, both choices are acceptable. The computation were repeated for sets of 200 trajectories with the initial epochs spaced 5 years apart over a period of 50 years. The propagations run for 130 years, in order to fully capture three disorder episodes (at approximately 5, 50 and 100 years after the initial epoch). The results are shown in Fig. 11, where the error bars indicate the start and end of the episodes, with the markers placed at the midpoint. The first disorder episode is denoted with a blue cross, the second with a red triangle and the third with a gray square The linear regression to the simulated data shows, as expected, a very strong correlation between the initial epoch and the occurrence of the transitions. The disorder episodes take place approximately 4.3, 54 and 104 years after the start of the propagation, which agrees very well with the length of the inclination cycle (53 years). Note that the transition at 4.3 years is actually a residual of an episode that would occur at 0 years, but it is suppressed because the initial condition forces the spacecraft to start tightly packed and with zero drift. Therefore, to determine the period between disorder episodes without contamination from the initial condition, one should consider the values 0, 54 and 104 years.

Keep in mind also that the episodes are not instantaneous, they span several years (~10), so there is uncertainty in the date. All considered, the concordance of the period between episodes with the theoretical inclination cycle is extremely good.

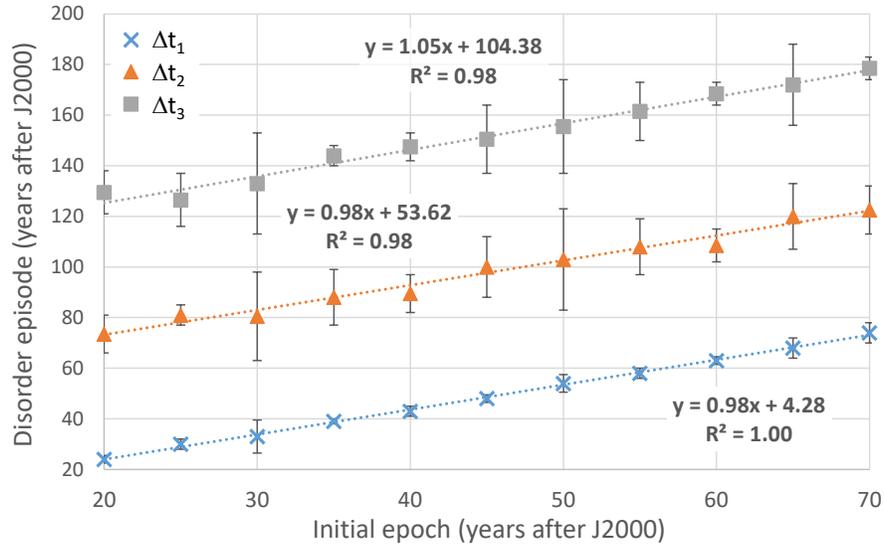

**Fig. 11    Time of the transitions vs. initial epoch (initial longitude 165.3ºE).**

A rough estimate of the strength of the transitions is obtained recording the magnitude of the largest jump in scatter that takes place during the transition (see $\Delta\sigma$ in Fig. 10). It was decided to measure a single jump in standard deviation instead of the total change across the transition because, when two consecutive direction reversals take place, they may partially cancel each other (one squeezes the left tail of the cloud, while the other compresses the right end). Thus, the net change can be smaller than the individual jumps. The authors believe that the magnitude of individual jumps gives a better indication of the strength of the transition. In any case, what really matters is the order of magnitude of the jumps. Exact values are not relevant because, due to the extreme sensitivity to initial conditions, different samples from the same initial distribution yield slightly different statistical parameters. As far as the order of magnitude is concerned, the net change and the individual jumps give comparable estimates. Thus, they are both acceptable strength measures. The strength vs. initial epoch plot is shown in Fig. 12. For reference, whenever a linear rate of increase of the standard deviation develops after the transition (e.g., in the baseline configuration) it has been recorded as "xx dpy", with "dpy" standing for degrees per year. Because the transition strength spans several orders of magnitude, the vertical axis uses a logarithmic scale. This has the added benefit of making the general appearance of the chart more robust. As indicated above, taking a different sample changes the jumps, typically by a factor less than 2 for a sample size of 200 spacecraft. Therefore, represented in logarithmic scale, the differences from sample to sample are limited. The strength of the first transition is relatively uniform, on the order of 1 degree (with all the values contained between 0.5 and 6.7 degrees). The second transition shows much higher variability (with jumps between 2.4 and 151 degrees). The highest strengths are for initial epochs around 2020, 2045 and 2070, giving rise to a subsequent linear drift in scatter. The third transition displays even higher variability, causing jumps between 2.8 and 327 degrees.

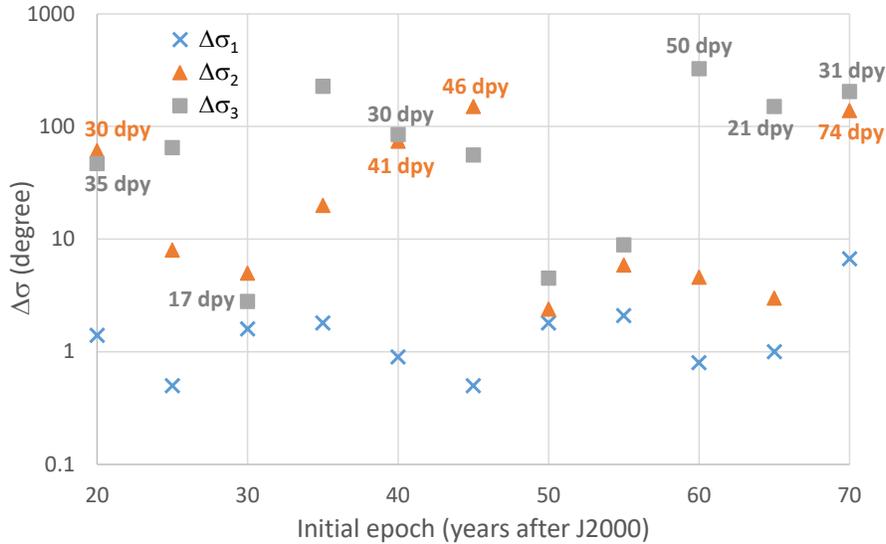

**Fig. 12    Strength of the transitions vs. initial epoch (initial longitude 165.3ºE).**

### V.    Simplified Modeling

As explained in a previous study [15], the most relevant harmonics of the gravity field are those of degree below 4. In particular, the degree-3 terms introduce the asymmetry between the two unstable equilibrium points that enables transitions between continuous circulation and long libration. Radiation pressure was found to play a minor role. Thus, a simplified physical model with only lunisolar perturbations and gravitational harmonics up to degree three was prepared. It also ignored Earth's precession and nutation effects for the sake of simplicity. This reduced model is expected to behave very close to the baseline setup.

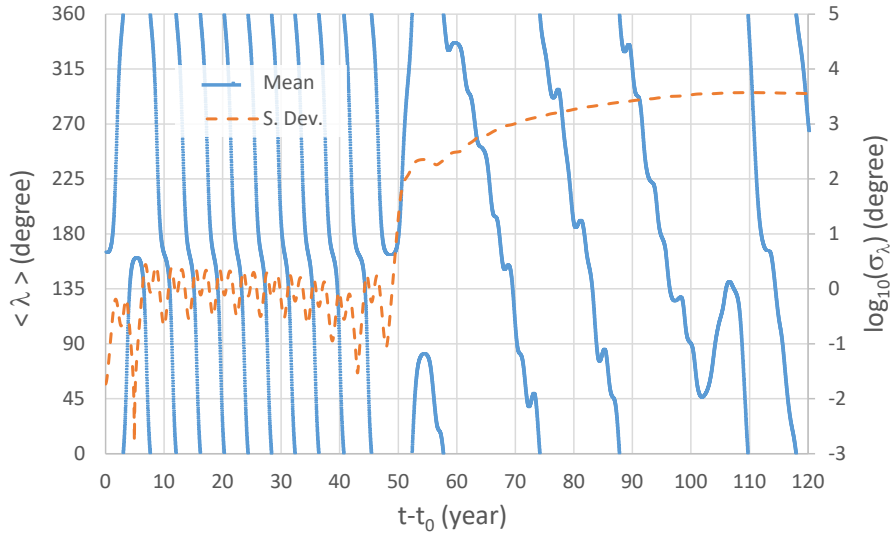

**Fig. 13    Longitude evolution for simplified model (N=3 + lunisolar perturbations).**

Fig. 13 demonstrates that this is indeed the case. The major features of the plot (drift reversals and jumps in scatter) coincide with the baseline solution (Fig. 1). Differences in mean longitude emerge after 60 years, but at that point the order of the cloud is already destroyed. Removing lunisolar perturbations (Fig. 14) makes the system autonomous in the Earth-fixed frame. Therefore, the transitions between modes of motion disappear leaving only continuous long libration without sudden increases in scatter. The cyclic variation of standard deviation due to passages across equilibrium points remains. On top, there is a slow secular trend, with the average standard deviation over one cycle increasing by less than 7° per century. This linear trend is just a reflection of the distribution of initial conditions, and does not involve any complex dynamical behavior.

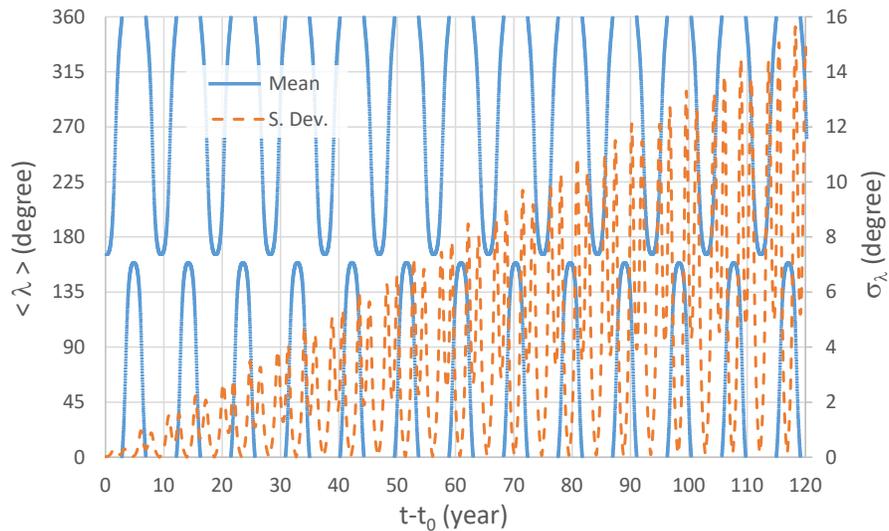

**Fig. 14    Longitude evolution for simplified model (N=3 only).**

Retaining lunisolar perturbations but only gravity harmonics of degree 2 makes the stable and unstable point pairs symmetric. This suppresses the long libration mode (see [15] for more details) forcing continuous circulation (Fig. 15). Due to the lunisolar perturbation, the potential barrier increases twice per century, causing a more pronounced deceleration of the cloud as it crosses the unstable positions. As explained before, the enhanced compression of the cloud results in increased scatter once the spacecraft accelerate. In this case, however, the increase is smooth because the motion is not reversed. The results in Fig. 14 and Fig. 15 indicate that, to preserve the qualitative behavior of the system, lunisolar perturbations and gravity harmonics up to degree 3 are required.

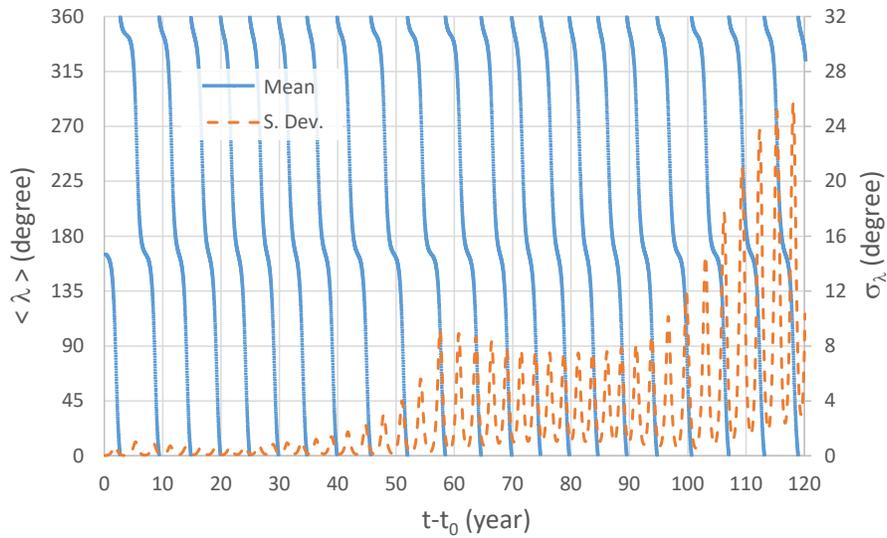

**Fig. 15** **Longitude evolution for simplified model (N=2 + lunisolar perturbations).**

### A. Simple Analytical Model for Earth and Moon Orbits

There is a good agreement between the time between transitions and the theoretical period of orbital precession. This suggests that, replacing precomputed ephemeris with a simple analytical model for the positions of Moon and Sun, should yield a reasonable approximation to the behavior of the system. It would also support the hypothesis that the orbital inclination cycle causes the sudden episodes of disorder. This was tested assuming coplanar (i.e., the lunar orbit is contained in the ecliptic) circular orbits for both Earth and Moon (with radii of $150 \cdot 10^6$ km and $385 \cdot 10^3$ km, respectively) and repeated the calculations. The physical model ignores radiation pressure and harmonics of the gravity field above degree 3. The results are presented in Fig. 16.

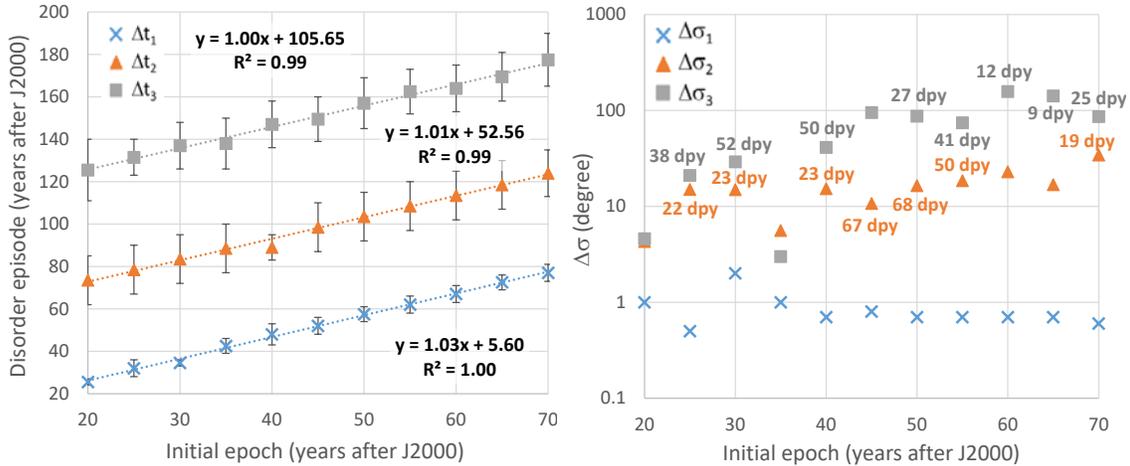

**Fig. 16** Transition time and strength vs. initial epoch for simple model.

As expected, the simple model is able to reproduce the timing of the changes in scatter very well. Also, the observed transition times agree even better with the theoretical cycle of precession, with the second and third episodes occurring 53 and 106 years after the initial epoch. The transition strength becomes more uniform, with the first jump remaining close to 1º while the second is on the order of 10º. Also, most of the transitions at the 50-year mark are followed by a linear increase in scatter. The third transition shows more variability but, as mentioned before, this data is harder to interpret because the changes occur after the cloud has become highly disordered.

**B. Refined Analytical Model for Earth and Moon Orbits**

Fig. 16 strongly suggests that the changes in transition strength are connected to the variability of the lunar orbit, neglected by the simple analytical model. This hypothesis was tested maintaining the circular orbits for Sun and Moon, but including the precession of the lunar orbital plane. Nodal precession is expected to be the most important factor, as it affects the inclination of the orbital plane of the Moon relative to the equator. Therefore, it changes the position of the pole of the Laplacian plane, which governs the inclination cycle of the spacecraft. The analytical model for the lunar orbit assumes a constant inclination relative to the ecliptic of 5.15º and a period of nodal precession of 18.6 years. The rest of the orbital parameters of the Moon were tuned to obtain the best match of the precomputed ephemeris on 1 January 2020. Like in the previous subsection, solar radiation pressure and gravity harmonics above degree 3 are not included. The results of this model are presented in Fig. 17, showing that the variability in strength is enhanced relative to the coplanar model. For reference, the right pane of Fig. 17 includes the inclination of the lunar orbit relative to the equator (dashed black line). The strongest transitions take place around 2020, 2045 and 2065; considering the simplicity of the model, the agreement with the reference solution (Fig. 11 and Fig. 12) is remarkable. Roughly speaking, cases starting when the inclination of the lunar orbit is highest tend to show stronger transitions. This aligns with the idea that the orbital inclination of the Moon plays an important role in the strength of the episodes.

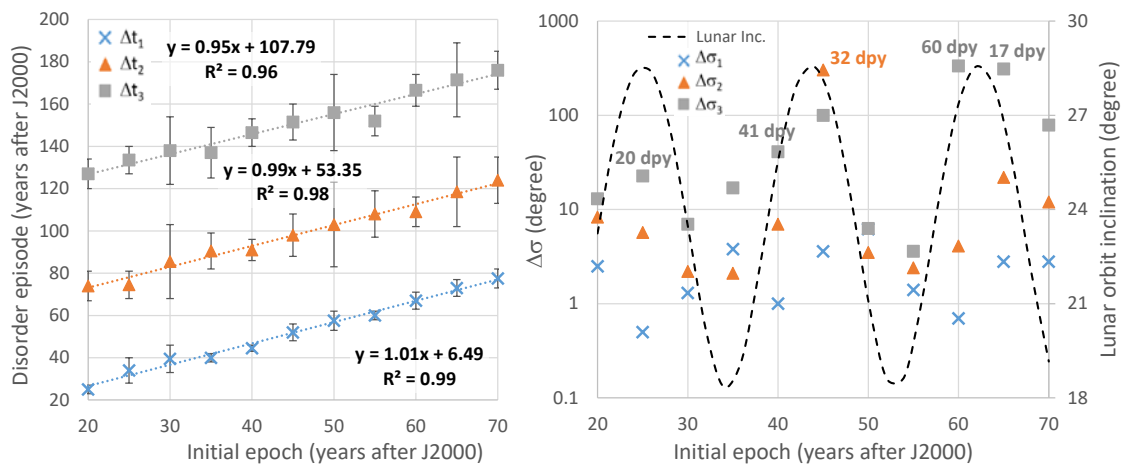

**Fig. 17** **Transition time and strength vs. initial epoch for enhanced simple model.**

## VI. Conclusions

The long-term evolution of geostationary spacecraft abandoned near the region of maximum instability of the geopotential (165ºE) was studied with a Monte Carlo simulation. The combined effect of lunisolar perturbations and harmonics of the gravity field up to degree 3 makes the system susceptible sudden episodes of disorder. These occur approximately every half century, at the points of minimum inclination of the precessional cycle of the orbital plane. The disorder is triggered by transitions between continuous circulation and long libration modes, which are extremely sensitive to perturbations, to the point of becoming effectively unpredictable.

While individual trajectories cannot be predicted accurately, the statistical behavior of the ensemble of spacecraft is significantly more robust. This is demonstrated by models of widely different levels of fidelity showing the same qualitative behavior. Therefore, the observed phenomena are easy to reproduce, because they depend weakly on the subtleties of the physical model.

The main conclusion is that the longitudinal motion of satellites drifting from the 165ºE position is unpredictable over scales of 40 years, requiring careful tracking to prevent accidental collisions. On a positive note, the unpredictability is only an issue when the orbital plane is close to the equator. For the rest of the spacecraft inclination cycle (which is itself extremely regular) the motion is easily predictable. This means only a small fraction of uncontrolled satellites require frequent updates of the trajectory analysis, those that at a given point in time have small inclinations (say, below 8º).

**Funding Sources**

The work of R. Flores and E. Fantino has been supported by Khalifa University of Science and Technology's internal grant CIRA-2021-65 / 8474000413. R. Flores also acknowledges financial support from the Spanish Ministry of Economy and Competitiveness "Severo Ochoa Programme for Centres of Excellence in R&D" (CEX2018-000797-S). In addition, E. Fantino received support from the Spanish Ministry of Science and Innovation under projects PID2020-112576GB-C21 and PID2021-123968NB-100.